\documentclass[epsfig,12pt,onecolumn]{article}
\usepackage{amsfonts}
\usepackage{amssymb}
\usepackage{amsmath}
\usepackage{multicol}
\usepackage{graphicx}
\usepackage{float}
\usepackage{caption}
\usepackage{subcaption}

\input{tcilatex}
\makeatletter \@addtoreset{equation}{section}

\def \be{\begin{equation}}
\def \ee{\end{equation}}
\def \bea{\begin{eqnarray}}
\def \eea{\end{eqnarray}}

\newcommand{\nc}{\newcommand}
\nc{\al}{\alpha} \nc{\bib}{\bibitem} \nc{\la}{\lambda}
\nc{\C}{\mbox{\hspace{1.24mm}\rule{0.2mm}{2.5mm}\hspace{-2.7mm} C}}
\nc{\R}{\mbox{\hspace{.04mm}\rule{0.2mm}{2.8mm}\hspace{-1.5mm} R}}

\begin{document}

\title{Interpretation of galaxy rotation curves from primordial black holes
in 4D Einstein-Gauss-Bonnet gravity}
\author{M. Bousder\thanks{%
mostafa.bousder@um5.ac.ma} \\
{\small LPHE-MS Laboratory, Department of physics,}\\
{\small Faculty of Science, Mohammed V University in Rabat, Morocco}}
\maketitle

\begin{abstract}
We develop a novel approach to the dark matter halos in the context of 4
dimensional Einstein--scalar-Gauss-Bonnet gravity to reproduce the flat
rotation curves of galaxies. Moreover, the Gauss-Bonnet coupling describes
the interior structure of the galaxies, while there is a presence of a
scalar field $\phi $ in the galaxy edges. This can provide an interesting
interpretation for the functional coupling $f(\phi )$. We discuss how this
comparison can naturally drive the observed percentages of matter and dark
matter in the Universe. The effective mass range\ in our model is $%
10^{-2}kg-10^{3}kg$, which is in good agreement with the constraints on
primordial black holes.

\textbf{Keywords:} Galaxy, rotation curves, primordial black holes,
Einstein-Gauss-Bonnet gravity
\end{abstract}

\section{ Introduction}

The first accepted evidence for the existence of dark matter in the galaxy
rotation curves \cite{VC1,VC2,VC3}. It is well known that the orbital
velocities $V$ of planets in planetary systems decline with distance
according to Kepler's third law \cite{80} $V^{2}=GM/r$. While the rotational
velocity of the galaxy almost stays consistently among all galaxies \cite{9,10}%
. The rotational speeds of stars inside the galaxy do not follow the rules
found in smaller orbital systems. The effective potential can be used to
determine the orbits of planets \cite{101}, also in the cosmological
evolution analysis of the scalar field, allowing the detection of dark
energy in orbit \cite{102}.\ In Modified Newtonian Dynamics (MOND), the
rotation curves of spiral galaxies are asymptotically flat \cite{MO1} and
imply a mass-velocity relationship as $V^{D}\propto M$, with $D$ is in the
neighborhood of $4$ \cite{MO01}. It is known that some interesting phenomena
of galaxies, like the relationship between the rotation velocity in spiral
galaxies and the luminosity \cite{MO2}. According to the MOND theory, the
rotational velocity of stars around a galaxy at large distances is $V_{\max
}^{4}\sim GMa_{0}$, where $a_{0}\approx 1,2\times 10^{-10}ms^{-2}$ is a
fundamental acceleration scale of nature, and $M$ is the total mass of a
galaxy. It is treated as a point mass at its center, providing a crude
approximation for a star in the outer regions of a galaxy. The $V_{\max }^{4}
$ predicts that the rotational velocity is constant out to an infinite range
and that the rotational velocity doesn't depend on a distance scale but the
magnitude of the acceleration $a_{0}$. Additionally, the amplitude and scale
of the initial fluctuations describe the formation of galactic halos in a
flat universe dominated by cold dark matter (CDM) \cite{DH1}. Massive halos
form preferentially in regions of high dark matter density. The formation of
dark halos is under the assumption that the CDM particles have a finite
cross section for elastic collisions \cite{DH2}. Their formation sites
correspond well to high peaks of the initial linear density field. In the
galactic nuclei, a possible explanation for the formation of the
supermassive black holes at the galaxy center is caused by the collapse of a
large number of stars' high concentrations at the galaxy center. To describe
the formation of protogalaxies, it's useful to use the second-order phase
transition in the inflation stage \cite{DH3}.\newline
Recently, the 4D Einstein-Gauss-Bonnet (EGB) theory \cite{N7} provides a new
insights into the 4-dimensional (4D) theory of gravity. It is in
contradiction with the Lovelock theorem \cite{R1} which describes the
gravity at $D\geq 5$. The idea of the 4D EGB gravity is rescaling first the
Gauss-Bonnet coupling constant by the factor $1/(D-4)$, then taking the
limit $D\longrightarrow 4$. The divergent factor $1/(D-4)$ is canceled by
the vanishing GB contributions, which leads to a theory of gravity with only
two dynamical degrees of freedom. However, the idea of the limit $%
D\longrightarrow 4$ is not clearly defined \cite{F1,F2,F3,F4}. It was
explicitly confirmed by a direct product $D$-dimensional spacetime or by
adding a counter term, before taking the limit $D\longrightarrow 4$, which
can be seen as a class of Horndeski theory \cite{H0}. Although the EGB
gravity is currently debatable, the spherically symmetric black hole
solution is still meaningful and worthy of study \cite{n}. In \cite{bsd} we
have studied the relationship between the MOND paradigm and
scalar-Gauss-Bonnet (EsGB) gravity, we added a new relativistic part to the
MOND from EsGB gravity.\newline
Motivated by these, the purpose of the present paper is to introduce a
difference between the physical quantities in the interior (ex: interior of
galaxies) and the edge (ex: dark matter halos). We will also point out that
there is the presence of two gravitational constants, the first is that of
Newton's constant, the second appears more for very intense gravity (like
galaxies). We will show why the values of the constant $\theta _{0}$ in \cite%
{bsd} for the dwarf spheroidal and irregular dwarf galaxies are not
constant. We show the presence of primordial black holes in the dark matter
halos according to the rotation curves of galaxies.

In section 2, we briefly review the Einstein-Gauss-Bonnet gravity in
4-dimensions in coupling with a scalar field. In Section 3, we study the
effective mass of this scalar field and its relation to the rotation curves
of galaxies. In Section 4, we study the galaxy formation from primordial
black holes. In Section 5, we summarize our conclusions.

\section{Dark halos in Einstein-Gauss-Bonnet gravity}

The EGB gravity is a higher derivative terms of the Lovelock gravity. In
this section, we explain in detail how to construct the equation of motion
of the Einstein-scalar-Gauss-Bonnet (EsGB) gravity. We begin by reviewing
the 4D EsGB action \cite{D1}%
\begin{equation}
\mathcal{S}=\frac{1}{2\kappa ^{2}}\int \sqrt{-g}d^{4}x\left( \mathcal{L}-%
\frac{1}{2}g^{\mu \nu }\partial _{\mu }\phi \partial _{\nu }\phi -\mathcal{V}%
\left( \phi \right) \right) +\mathcal{S}_{m}.  \label{F1}
\end{equation}%
The EsGB lagrangian is given $\mathcal{L}=R+f\left( \phi \right) \left(
R^{2}-4R_{\mu \nu }R^{\mu \nu }+R_{\mu \nu \rho \sigma }R^{\mu \nu \rho
\sigma }\right) ,$ where $1/\kappa ^{2}=1/8\pi G_{N}=1.221\times 10^{19}GeV$
is the reduced Planck mass, $R$ is the Ricci scalar, $\mathcal{S}_{m}$ is
the matter action and $f\left( \phi \right) $ is a functional coupling of
the scalar field $\phi $. In the above equation $\left( \mu ,\nu \right)
=\left( 0,1,2,3\right) $. The variation with respect to the field $\phi $
gives us the equation of motion for the scalar field%
\begin{equation}
\square \phi =\partial _{\phi }\mathcal{V}_{eff}\left( \phi \right) ,
\label{F3}
\end{equation}%
where $\square \equiv \nabla _{\mu }\nabla ^{\mu }$ and the effective
potential is%
\begin{equation}
\mathcal{V}_{eff}\left( \phi \right) =\mathcal{V}\left( \phi \right)
-f\left( \phi \right) \mathcal{G}.  \label{F4}
\end{equation}%
Varying the action (\ref{F1}) over the metric $g_{\mu \nu }$, we obtain the
following equations of motion:%
\begin{equation}
G^{\mu \nu }+\mathcal{K}^{\mu \nu }+f\left( \phi \right) \mathcal{H}^{\mu
\nu }+\frac{1}{2}\left[ \mathcal{T}_{\phi }^{\mu \nu }-g^{\mu \nu }\mathcal{V%
}_{eff}\left( \phi \right) \right] =\frac{1}{2}\kappa ^{2}T^{\mu \nu },
\label{EM}
\end{equation}%
where the Einstein tensor is $G^{\mu \nu }=R^{\mu \nu }-\frac{1}{2}g^{\mu
\nu }R$, the matter stress tensor is $T^{\mu \nu }=-\frac{2}{\sqrt{-g}}\frac{%
\delta \mathcal{S}_{m}}{\delta g_{\mu \nu }}$. On the other hand, the $%
\mathcal{K}^{\mu \nu }$ and $\mathcal{H}^{\mu \nu }$ are given by%
\begin{equation}
\mathcal{K}^{\mu \nu }=4\left[ G^{\mu \nu }\square +\frac{1}{2}R\nabla ^{\mu
}\nabla ^{\nu }+\left( g^{\mu \nu }R^{\rho \sigma }-R^{\mu \rho \nu \sigma
}\right) \nabla _{\rho }\nabla _{\sigma }-R^{\nu \rho }\nabla _{\rho }\nabla
^{\mu }+R^{\mu \rho }\nabla _{\rho }\nabla ^{\nu }\right] f\left( \phi
\right) ,  \label{F6}
\end{equation}%
\begin{equation}
\mathcal{H}^{\mu \nu }=2R^{\mu \rho \sigma \tau }R_{\text{ \ }\rho \sigma
\tau }^{\nu }-RR^{\mu \nu }+\frac{1}{2}R_{\text{ \ }\rho }^{\mu }R^{\nu \rho
}-R^{\mu \rho \sigma \tau }R_{\text{ \ }\rho \sigma \tau }^{\nu }.
\label{F7}
\end{equation}%
The tensor $\mathcal{K}^{\mu \nu }$ represents an operator which acts on $%
f\left( \phi \right) $. The energy-momentum tensor for the scalar field is%
\begin{equation}
\mathcal{T}_{\phi }^{\mu \nu }=\nabla ^{\mu }\phi \nabla ^{\nu }\phi -\frac{1%
}{2}g^{\mu \nu }\nabla _{\rho }\phi \nabla ^{\rho }\phi .  \label{F8}
\end{equation}%
The form the functional $f\left( \phi \right) $ can take $f\left( \phi
\right) \propto e^{-\gamma \phi }$ \cite{L1}, where $\gamma $ is a constant,
which corresponds to EGB gravity coupled with dilaton that arises as a
low-energy limit of the string theory \cite{L2}. The motion of compound
objects in an external field\ of the galaxy like the globular clusters is
independent of its internal structure and may be described in the MOND limit
\cite{BM}. We assume that in the galaxies\ edges (dark halos), there is a
presence of a scalar field $\phi $, while inside the galaxy is replaced by
the Gauss-Bonnet (GB) coupling $\alpha $ as
\begin{equation}
\begin{array}{c}
\text{coupling constant=}f\left( \phi \right) ,\text{ \ \ galaxy\ edges }%
\left( D=4\right)  \\
\text{coupling constant=}\frac{\alpha }{D-4},\text{ \ \ galaxy\ interior }%
\left( D\rightarrow 4\right)
\end{array}%
.  \label{F11}
\end{equation}%
The GB coupling $\alpha $ is measured in\textbf{\ }$km^{2}$. In the
galaxies\ interior, we have rescaled the coupling constant $\alpha
\rightarrow \alpha /\left( D-4\right) $. The negative (positive) $\alpha $
leads to a decrease (increase) of the galaxy radius and the maximum mass
\cite{L5}. If $\alpha <0$\ the solution is still the anti-de Sitter (AdS)
space, if $\alpha >0$\ the solution is the de Sitter (dS) space \cite{L3}.%
\newline
We investigate in detail the impact of the Gauss--Bonnet coupling on
properties of the galaxies, such as mass, radius and density. Considering
the limit $D\rightarrow 4$, it has an effect on gravitational dynamics in
4D. Additionally, at the galaxy boundary $(r=r_{\max })$, the GB coupling
must be continuous, i.e. $f\left( \phi \right) \rightarrow \frac{\alpha }{D-4%
}$. On the other hand, the function $f\left( \phi \right) $ describes the
star exterior region. To study the equations of motion inside and outside
the galaxy, we differentiate between two cases:\newline
In galaxy\ interior ($D\rightarrow 4$) we have:%
\begin{equation}
G^{\mu \nu }+\alpha \left( \mathcal{H}^{\mu \nu }+\frac{1}{2}g^{\mu \nu }%
\mathcal{G}\right) =\frac{\kappa ^{2}}{2}T^{\mu \nu }.  \label{F12}
\end{equation}%
In galaxy\ edges ($D=4$) we have:%
\begin{equation}
G^{\mu \nu }+\mathcal{K}^{\mu \nu }+f\left( \phi \right) \left[ \mathcal{H}%
^{\mu \nu }-\frac{1}{2f\left( \phi \right) }g^{\mu \nu }\left( \nabla
_{\lambda }\phi \nabla ^{\lambda }\phi +4\mathcal{V}_{eff}\left( \phi
\right) \right) \right] =\frac{1}{2}\kappa ^{2}T^{\mu \nu },  \label{F13}
\end{equation}%
with $g_{\mu \nu }\mathcal{T}_{\phi }^{\mu \nu }=-\nabla _{\lambda }\phi
\nabla ^{\lambda }\phi $. In galaxy\ interior ($D\rightarrow 4$) we have:\
the GB invariant can be greatly simplified to the matter density \cite{DM,OV}%
. By comparing, Eq.(\ref{F12}) and Eq.(\ref{F13}), we notice that the term $%
\left( \nabla _{\lambda }\phi \nabla ^{\lambda }\phi +4\mathcal{V}%
_{eff}\left( \phi \right) \right) /f\left( \phi \right) $ represents a
density. Using Eq.(\ref{F4}) we obtain%
\begin{equation}
\begin{array}{c}
\rho _{DM}\equiv 4\mathcal{G}-\frac{4}{f\left( \phi \right) }\left( \frac{1}{%
4}\nabla _{\lambda }\phi \nabla ^{\lambda }\phi +\mathcal{V}\left( \phi
\right) \right) ,\text{ \ \ galaxy\ edges }\left( D=4\right)  \\
\rho _{m}=\mathcal{G},\text{ \ \ \ \ \ \ \ \ \ \ \ \ \ \ \ \ \ \ \ \ \ \ \ \
\ \ \ \ \ \ \ \ \ \ \ \ \ \ galaxy\ interior }\left( D\rightarrow 4\right)
\end{array}%
.  \label{F14}
\end{equation}%
where $\rho _{m}\left( r\right) $ is the density of matter enclosed within $r
$, and $\rho _{DM}$ is the density of dark matter halo surrounding the
galaxy. Note that the relation between $\rho _{DM}$ and $\rho _{m}$
highlight the chameleon dark matter \cite{L12,L13}. For $\mathcal{V}\left(
\phi \right) \approx -\frac{1}{4}\nabla _{\lambda }\phi \nabla ^{\lambda
}\phi $, we obtain $\rho _{DM}\approx 4\rho _{m}$, which is in good
agreement with the observation data of the percentages of dark matter and
the matter in the Universe \cite{L16}: $\rho _{DM}\equiv 80\%$ and $\rho
_{m}\equiv 20\%$. In this profile, the density $4\rho _{m}-\rho _{DM}$
represents the small variation of the DM density according to the dynamics
of the field $\phi $.

\section{Rotation curves of galaxies}

The scalar field sits at the minimum of its effective potential. We assume
that a massive scalar field begins oscillating about a minimum. The mass of
small fluctuations around $\phi _{\min }$ gives a new scalar field mass as
effective mass by
\begin{equation}
M_{eff}^{2}=\left. \frac{\partial ^{2}}{\partial \phi ^{2}}\mathcal{V}%
_{eff}(\phi )\right\vert _{\phi =\phi _{\min }}.  \label{r2}
\end{equation}%
The effective mass is expressed in the following way%
\begin{equation}
M_{eff}=\sqrt{\frac{\partial ^{2}\mathcal{V}\left( \phi _{\min }\right) }{%
\partial \phi ^{2}}-\rho _{m}\frac{\partial ^{2}f\left( \phi _{\min }\right)
}{\partial \phi ^{2}}}.  \label{r3}
\end{equation}%
This equation is in good agreement with the expression for the effective
mass in \cite{BC,BC1}. At the end of inflation scenario \cite{BC}, the mass $%
M_{eff}$ is described by sinusoidal functions by the quantum fluctuations.
During each oscillation of the field $\phi $, the effective mass $M_{eff}$
is much greater than the inflaton mass $\sqrt{\rho _{m}\partial ^{2}f\left(
\phi _{\min }\right) /\partial \phi ^{2}}$. Therefore, the resonance of $%
\phi $ begins at the end of inflation with the typical frequency of
oscillation $\omega (t)\sim \sqrt{M_{eff}^{2}(t)+\rho _{m}\partial
^{2}f\left( \phi _{\min }\right) /\partial \phi ^{2}}$. For very small $\phi
$ there is a change in the frequency of oscillations $\omega (t)$ and the
system becomes adiabatic. From Eq. (\ref{r3}), we notice the existence of
the condition:%
\begin{equation}
\rho _{m}\leq \frac{\mathcal{V}^{\prime \prime }\left( \phi _{\min }\right)
}{f^{\prime \prime }\left( \phi _{\min }\right) },  \label{r4}
\end{equation}%
with ($\prime $) represent $\partial /\partial \phi $. Taking $\rho _{\max }=%
\mathcal{V}^{\prime \prime }\left( \phi _{\min }\right) /f^{\prime \prime
}\left( \phi _{\min }\right) $, with $\rho _{\max }$ represents the maximum
mass of ordinary matter in the galaxy. Using Eqs. (\ref{r3},\ref{r4}) we find%
\begin{equation}
M_{eff}=\sqrt{\frac{\partial ^{2}f\left( \phi _{\min }\right) }{\partial
\phi ^{2}}}\sqrt{\rho _{\max }-\rho _{m}\left( r\right) },  \label{r5}
\end{equation}%
where $\rho _{m}\left( r\right) $ is the local matter energy density and $%
f\left( \phi _{\min }\right) $. We notice that $\rho _{\max }-\rho
_{m}\left( r\right) \sim M_{eff}^{2}\sim $(rotation velocity)$^{4}$. For the
elliptical galaxy, we use the power-law relation between the luminosity $L$
and the central stellar velocity dispersion $\sigma $ of Faber-Jackson
relation: $L\propto \sigma ^{D}$, where $D$ is in the neighborhood of $4$
and $\sigma $ is the stellar velocity dispersion. While for the spiral
galaxy we use the Tully--Fisher relation (TFR), it is a relationship between
the mass (intrinsic luminosity) and its asymptotic rotation velocity. The
gravitational potential is writing as $\nabla ^{2}V=4\pi G\rho _{m}$ \cite%
{MO3}. Using the mass-velocity relationship as is shown in \cite{bsd}, we
notice that$\sqrt{\partial ^{2}f\left( \phi _{\min }\right) /\partial \phi
^{2}}=GM/KV^{2}$, where $V$ is the rotation velocity of the galaxy disk and $%
K$ is a constant, which roughly equals $K\approx 69.44kg^{-1/2}km^{-1/2}$
and\ $\theta _{0}=\left( KM_{eff}\right) ^{-1/2}$ \cite{bsd}. The total mass
of the galaxy $M$ is integrated mass within some the galaxy radius $r_{\max }
$ as $M=\int_{0}^{r_{\max }}4\pi r^{2}\rho _{m}\left( r\right) dr$. Notice
that we have $\rho _{m}\left( r_{\max }\right) \neq \rho _{\max }$. It is
clear from the expression in Eq. (\ref{r5}) that
\begin{equation}
M_{eff}V^{2}=\frac{GM}{K}\sqrt{\rho _{\max }-\rho _{m}\left( r\right) }.
\label{r6}
\end{equation}%
This expression of $M_{eff}$ is in good agreement with the masses of the
produced PBHs following the critical scaling in \cite{NANO4,NANO5}. The
effective mass $M_{eff}$ may represent the mass of the primordial black
hole. To verify this, we calculate the masses $M_{eff}$ for some galaxies;
see tables (\ref{table1}) and (\ref{table1}). We found that this mass varied
in the interval $M_{eff}\sim 10^{-2}kg-10^{3}kg$, which is in good agreement
with the constraints on the fraction of the Universe that may have gone into
PBHs over the mass range $10^{-8}kg-10^{47}kg$ \cite{PB10}. We note that the
term $M_{eff}V^{2}/2$ is the kinetic energy of field $\phi $. Additionally,
since $V$ is the galaxy rotation velocity, then this field is responsible
for the galaxies rotation. In the limit of $KM_{eff}=r\sqrt{\rho _{\max
}-\rho _{m}\left( r\right) }$ we can only recover the corresponding Poisson
equation for Newtonian velocity $V^{2}=GM/r$. In the galaxy\ edges, the
local density of ordinary matter $\rho _{m}\left( r\right) $ is very low,
from which, we obtain%
\begin{equation}
V_{\max }=\left( \frac{GM}{KM_{eff}}\right) ^{1/2}\rho _{\max }^{1/4},
\label{r7}
\end{equation}%
where\ $V_{\max }$\ is the maximum rotation velocity of the galaxy.
According to this relation, the Milgrom constant $a_{0}$ is no longer
constant, but it depends on other parameters like the mass of the galaxy and
the effective mass. If $M_{eff}$ is nearly constant, the $V_{\max }$ is
completely flat. There are two gravitational constants $(G,K)$. This means
that in the limit of low mass (sun, stars, planets), the $KM_{eff}$ term
will be very weak. For the galaxy and galaxy clusters masses, the term $%
KM_{eff}$ will be more important and will have an impact on the rotation
curves. It is possible that DM does not consist of a particle, but we just
must to find a theory of gravity beyond general relativity \cite{OV}.
\begin{table}[H]
\begin{equation*}
\begin{tabular}{cccccc}
\hline\hline
LS galaxies & $\rho (\times 10^{-22}kg/m^{2})$ & $M_{in}(\times
10^{10}M_{\odot })$ & $V_{\max }(km/s)$ & $\theta _{0}\left(
kg^{-1/4}km^{1/4}\right) $ & $M_{eff}(kg)$ \\ \hline\hline
Milky Way & $13,10$ & $12,10$ & $220$ & $0,086$ & $1.94$ \\
NGC 7331 & $8,70$ & $14,70$ & $268.1$ & $0,105$ & $1.30$ \\
NGC 4826 & $45,00$ & $1,90$ & $180.2$ & $0,130$ & $0.85$ \\
NGC 6503 & $18,40$ & $0,958$ & $121$ & $0,154$ & $0.61$ \\
NGC 7793 & $12,00$ & $0,88$ & $117.9$ & $0,174$ & $0.47$ \\
UGC 2885 & $15,30$ & $11,70$ & $300$ & $0,114$ & $1.11$ \\
NGC 253 & $10,00$ & $4,30$ & $229$ & $0,160$ & $0.56$ \\
NGC 925 & $8,60$ & $2,00$ & $113$ & $0,120$ & $1$ \\
NGC 2403 & $5,20$ & $2,90$ & $143.9$ & $0,144$ & $0.69$ \\
NGC 2841 & $8,02$ & $17$ & $326$ & $0,121$ & $0.98$ \\
NGC 2903 & $6,30$ & $6,70$ & $215.5$ & $0,135$ & $0.79$ \\
NGC 3198 & $3,26$ & $6,00$ & $160$ & $0,125$ & $0.921$ \\
NGC 5585 & $10,1$ & $0,59$ & $92$ & $0,173$ & $0.48$ \\
NGC 4321 & $8,80$ & $16,8$ & $270$ & $0,098$ & $1.49$ \\ \hline\hline
MS galaxies & $\rho (\times 10^{-22}kg/m^{2})$ & $M_{in}(\times
10^{10}M_{\odot })$ & $V_{\max }(km/s)$ & $\theta _{0}\left(
kg^{-1/4}km^{1/4}\right) $ & $M_{eff}(kg)$ \\ \hline\hline
NGC 4303 & $6,81$ & $3,68$ & $150$ & $0,124$ & $0.93$ \\
NGC 5055 & $5,21$ & $7,07$ & $215$ & $0,138$ & $0.75$ \\
NGC 4736 & $14,00$ & $1,77$ & $198.3$ & $0,198$ & $0.36$ \\
NGC 5194 & $1,00$ & $4,00$ & $232$ & $0,299$ & $0.16$ \\
NGC 4548 & $3,20$ & $3,80$ & $290$ & $0,287$ & $0.17$ \\ \hline\hline
L and E Galaxies & $\rho (\times 10^{-22}kg/m^{2})$ & $M_{in}(\times
10^{10}M_{\odot })$ & $V_{\max }(km/s)$ & $\theta _{0}\left(
kg^{-1/4}km^{1/4}\right) $ & $M_{eff}(kg)$ \\ \hline\hline
UGC 3993 & $3,10$ & $17.8$ & $300$ & $0,138$ & $0.75$ \\
NGC 7286 & $4,60$ & $0.59$ & $98$ & $0,224$ & $0.28$ \\
NGC 2768 & $10,00$ & $1.98$ & $260$ & $0,268$ & $0.20$ \\
NGC 3379 & $0,90$ & $1.10$ & $60$ & $0,151$ & $0.63$ \\
NGC 2434 & $1,00$ & $5.00$ & $231$ & $0,266$ & $0.20$ \\
NGC 4431 & $13,00$ & $0.30$ & $78$ & $0,193$ & $0.38$ \\ \hline
\end{tabular}%
\end{equation*}%
\caption{Large spirals (LS) galaxies, Messier Spirals (MS), Lenticular and
Elliptical (L and E) Galaxies.}
\label{table1}
\end{table}

\begin{figure}[H]
\centering
\includegraphics[width=12cm]{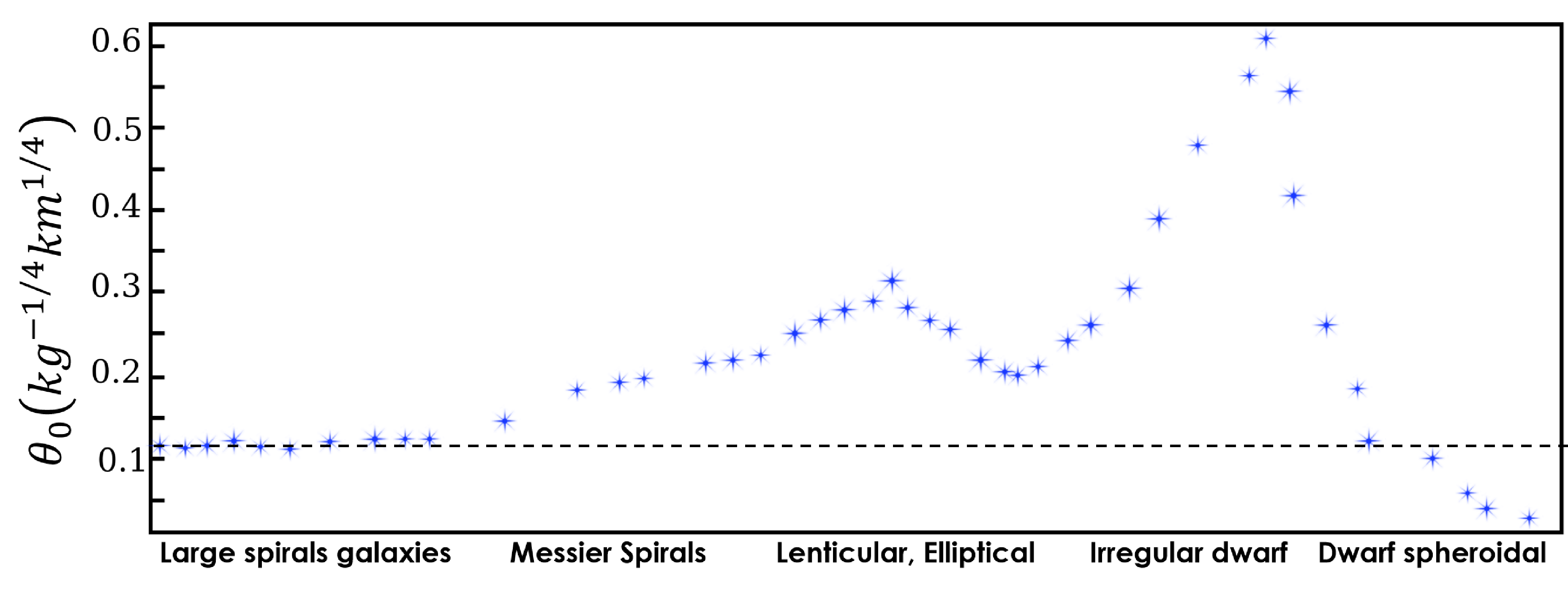}
\caption{Evolution of $\protect\theta _{0}=\left( KM_{eff}\right) ^{-1/2}$
as a function of type of galaxies. }
\label{f1}
\end{figure}

\begin{figure}[H]
\centering
\includegraphics[width=12cm]{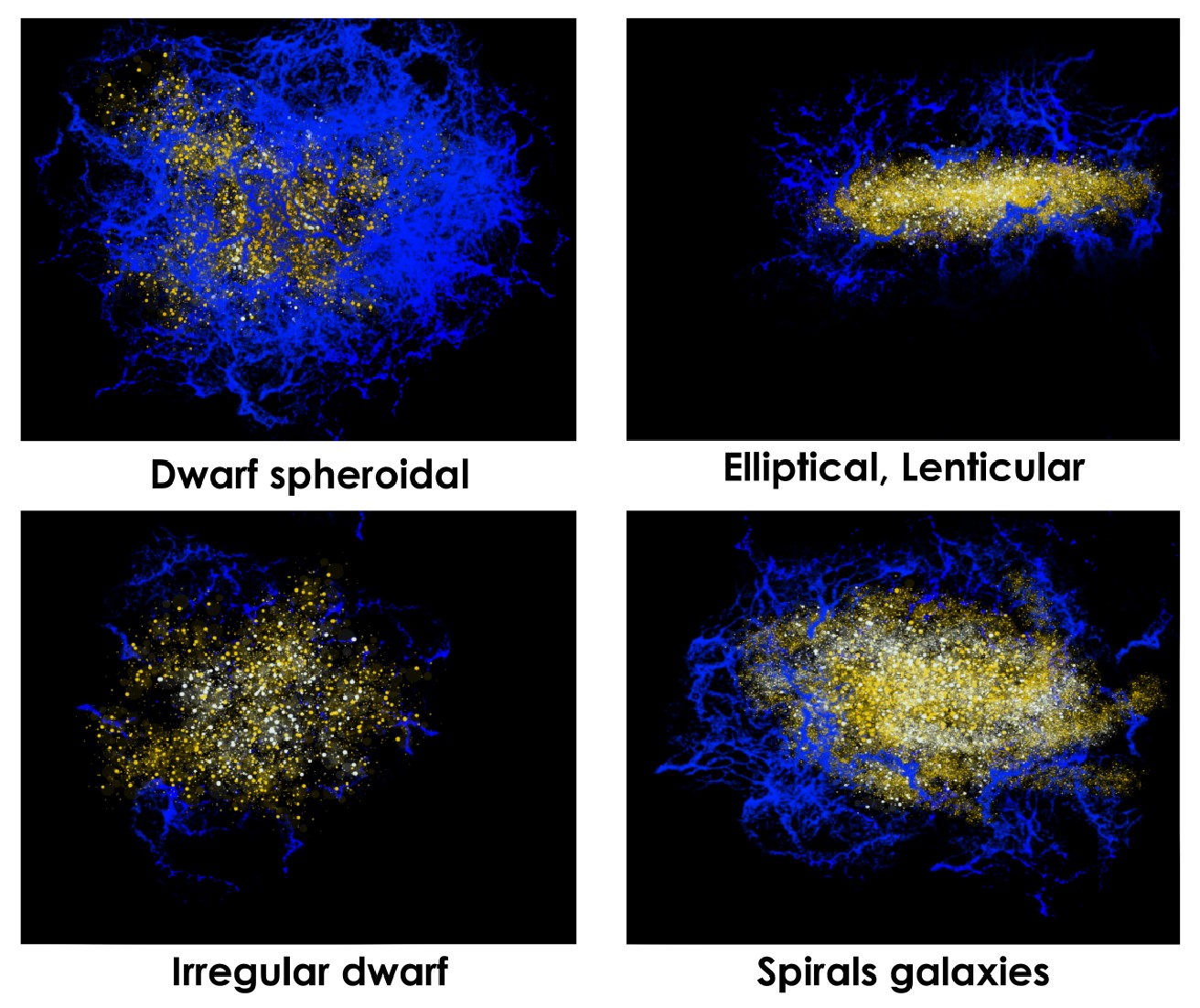}
\caption{A representation of the evolution of $M_{eff}$ as a function of
type of galaxies. The dark matter halo is described by the value of $M_{eff}$%
. The blue color represents the halo of dark matter and the white color
represents the matter of the galaxies.}
\label{f2}
\end{figure}

\begin{table}[H]
\begin{equation*}
\begin{tabular}{cccccc}
\hline
Id galaxies & $\rho (\times 10^{-22}kg/m^{2})$ & $M_{in}(\times
10^{10}M_{\odot }) $ & $V_{\max }(km/s)$ & $\theta _{0}\left(
kg^{-1/4}km^{1/4}\right) $ & $M_{eff}(kg)$ \\ \hline\hline
WLM (DDO 221) & $0,92$ & $0,00863$ & $19$ & $0,539$ & $0.049$ \\
M81dWb & $5,00$ & $0,007$ & $28,5$ & $0,588$ & $0.041$ \\
Holmberg II & $3,64$ & $0,0428$ & $34$ & $0,307$ & $0.152$ \\
NGC 3109 & $8,00$ & $0,0299$ & $67$ & $0,605$ & $0.039$ \\
NGC 4789a & $93,00$ & $0,0188$ & $50$ & $0,303$ & $0.156$ \\
NGC 3034 & $22,00$ & $1,00$ & $137$ & $0,163$ & $0.542$ \\ \hline\hline
dSphs Galaxies & $\rho (\times 10^{-22}kg/m^{2})$ & $M_{in}(M_{\odot })$ & $%
V_{\max }(km/s)$ & $\theta _{0}\left( kg^{-1/4}km^{1/4}\right) $ & $%
M_{eff}(kg)$ \\ \hline\hline
Carina & $6,50$ & $3.38\times 10^{6}$ & $8,5$ & $0,007$ & $293.89$ \\
Leo I & $13,60$ & $7.74\times 10^{6}$ & $12,5$ & $0,006$ & $400.02$ \\
Draco & $7,40$ & $3.40\times 10^{6}$ & $12$ & $0,01$ & $144.00$ \\
Fornax & $0,373$ & $12.40\times 10^{6}$ & $11,5$ & $0,01$ & $144.00$ \\
\hline
\end{tabular}%
\end{equation*}%
\caption{Irregular dwarf (Id) galaxies and dwarf spheroidal (dSphs) galaxies}
\label{table2}
\end{table}
By comparing the two tables \ref{table1}\ and \ref{table2}, we notice that
the mass $M_{eff}$ is almost constant and small for the spirals galaxies,
while, it is very important for the dwarf spheroidal galaxies (dSphs) \cite%
{PB8}. This shows why the rotation curves of the spirals galaxies are almost
static. However, the curves of dSphs increase relatively to the galactic
center \cite{PB7}, which justifies the major presence of the dark matter in
dSphs \cite{PB9}. So, there is an interesting connection between the
parameters $M_{eff}$ and the amount of dark matter in each galaxy. This adds
a new physical parameter to Eq. (\ref{r7}), i.e. the mass $M_{eff}$ is among
the physical parameters of the galaxy in this model. $M_{eff}$ varied from
one type of galaxy to another, which shows that $M_{eff}$ is a parameter
that can determine the type of a galaxy if we know his rotational velocity.
Also, permit us to determine the rotational velocity if we know the type of
galaxy. From the observation data of galaxies, we trace the evolution of $%
\theta _{0}$\ and $M_{eff}$ according to the types of galaxies Fig. (\ref{f1}%
) and Fig. (\ref{f2}).
According to Fig. (\ref{f1}), the mass parameter $M_{eff}$ describes the
evolution and formation of galaxies. The $KM_{eff}$ in Fig. (\ref{f1})
should deviate systematically from $K\approx 69.44kg^{-1/2}km^{-1/2}$
according to the variation of $M_{eff}$. This shows that the parameter $%
M_{eff}$ plays an important role in describing both the rotation and the
type of galaxies. The evolution of $M_{eff}$ is done according to categories
of galaxies. This shows that dark matter changes from one type to another
type of galaxies. We introduce the effective density
\begin{equation}
\rho _{eff}\left( r\right) =\rho _{\max }-\rho _{m}\left( r\right) =\left(
\frac{KM_{eff}}{r}\right) ^{4}.  \label{r8}
\end{equation}%
The effective density $\rho _{eff}\sim M_{eff}^{4}$ is of order $4$, which
corresponds to the Standard Model (SM) radiation energy density: $\rho _{R}=%
\frac{\pi ^{2}}{30}g_{\ast }T^{4}$, where $g_{\ast }$ denotes the effective
number of relativistic degrees of freedom at reheating time. In this case,
the density describes the DM-gas interaction with a typical emission
temperature. During the radiation dominated era, the evolution of $\rho
_{eff}$ can thus lead to PBH production the density, see next section. As is
shown in \cite{PB4}, the radiation density depends on scale factor as $\rho
_{R}\propto a^{-4}$ in the era of radiation dominance, while the PBH density
checked $\rho _{PBH}\propto a^{-3}$. So, $M_{eff}$ depends on the redshift $%
z $ as $M_{eff}\propto a=1/\left( 1+z\right) $. This shows a link between
dark matter ($M_{eff}$) and the redshift of a halo which is in good
agreement as is shown in \cite{PB6}.

\section{Galaxy formation from primordial black holes}

In cosmological and quintessence behavior \cite{L14,L15}, the energy density
$\rho _{\phi }$ and pressure $P_{\phi }$ of the scalar field are given by%
\begin{equation}
\rho _{\phi }=\frac{1}{2}\dot{\phi}^{2}+\mathcal{V}\left( \phi \right) ,%
\text{ }P_{\phi }=\frac{1}{2}\dot{\phi}^{2}-\mathcal{V}\left( \phi \right) .
\label{F15}
\end{equation}%
The quintessence models describe the dark energy with a scalar field $\phi $%
. In this case, $\rho _{\phi }$ and $P_{\phi }$ are respectively, the
density and the pressure of the dark energy (DE). Next, we assume that $\phi
=\phi (t)$, i.e. $\nabla _{\lambda }\phi \nabla ^{\lambda }\phi =\dot{\phi}%
^{2}$. Starting from Eqs. (\ref{F14},\ref{F15}), we obtain, the density of
dark matter profile:%
\begin{equation}
\rho _{DM}=4\rho _{m}+\frac{1}{f\left( \phi \right) }\left( P_{\phi }-3\rho
_{\phi }\right) .  \label{F16}
\end{equation}%
For small $f\left( \phi \right) $, the DM density depends only on $\left(
P_{\phi }-3\rho _{\phi }\right) /f\left( \phi \right) $. To study the
stability of the DM\textbf{\ }under scalar field perturbations, we employ
the effective sound speeds $c_{\phi }^{2}=\delta P_{\phi }/\delta \rho
_{\phi }$. The effective sound speeds are related to the energy density and
the pressure, which checked the causality condition $0\leq c_{\phi }^{2}\leq
c^{2}$. From Eqs. (\ref{F4},\ref{F15},\ref{F16}) we obtain%
\begin{equation}
M_{eff}^{2}=\frac{1}{2}\frac{\partial ^{2}}{\partial \phi ^{2}}\left( \rho
_{\phi }-P_{\phi }-\frac{\rho _{m}}{\rho _{DM}-4\rho _{m}}\left( P_{\phi
}-3\rho _{\phi }\right) \right) _{\phi =\phi _{\min }}.  \label{F17}
\end{equation}%
From Eq. (\ref{F16}), the functional coupling is given by the dimensionless
fraction%
\begin{equation}
f\left( \phi \right) =\frac{P_{\phi }-3\rho _{\phi }}{\rho _{DM}-4\rho _{m}}.
\label{F18}
\end{equation}%
In the galaxy interior, we assume that $\rho _{DM}\approx 0$ and $P_{\phi
}\approx 0$, so we get $f\left( \phi \right) \approx 3\rho _{\phi }/4\rho
_{m}$. Since $\rho _{\phi }$ represents the density of DE according to
quintessence, the effect of DE is weak in the galaxy\ edges, which shows
that $f\left( \phi \right) \left( \propto \rho _{\phi }\right) \rightarrow 0$%
. Since the field $\phi $ exists in the galaxy edges, so, inside the galaxy,
we have ($P_{\phi }=\rho _{\phi }=0$), i.e. $f\left( \phi \right) =0$, which
exactly corresponds with the assumption Eq. (\ref{F11}). For this reason, we
exclude $f\left( \phi \right) $ inside galaxies, and we replace it with the
GB coupling $\alpha $. Note the existence of two branches for the dark halo
(DH) fraction $f_{DH}\left( \phi \right) $ in the galaxy\ edges ($\rho
_{m}\approx 0$ \cite{PB5}), depending on the sign chosen as
\begin{equation}
f_{DH}\left( \phi \right) =\left( \omega _{\phi }-3\right) \frac{\Omega
_{\phi }}{\Omega _{DM}},  \label{F19}
\end{equation}%
where $\Omega _{\phi }=\rho _{\phi }/\rho _{crit}$ and $\Omega _{DM}=\rho
_{DM}/\rho _{crit}=0.26$ \cite{L16} is the current DM density parameter. In
the context of dark energy (DE), the scalar field requiring $\omega _{\phi
}=P_{\phi }/\rho _{\phi }\approx -1$. The Planck Collaboration results \cite%
{L16} provides a constraints on the DE equation of state $\omega _{\phi
}\approx -1.028\pm 0.032$, i.e. $f_{halos}\left( \phi \right) \leq 0$. Since
the field $\phi $ represents the DM halos, then this DE-$\omega _{\phi }$
proposition is not valid. We recall that the Eq. (\ref{r3}) is in good
agreement with the expression for the effective mass in \cite{BC,BC1}, which
generates a relationship between the primordial black holes (PBHs) and DM.
It is argued that the PBHs could be the origin of the dark matter halos \cite%
{PB2,PB3}. The PBHs might form a considerable fraction of the DM (contribute
more than $10\%$ of the dark matter) \cite{PB1,BC1}, the field $\phi $ may
describes the fraction of PBHs. We recall that the PBH fraction defined as $%
f_{PBH}=\Omega _{PBH}/\Omega _{DM}<1$, where $\Omega _{PBH}$ is the PBH
abundance. We assume that $\Omega _{\phi }=\Omega _{PBH}$, yield
\begin{equation}
f_{PBH}=\frac{f_{DH}\left( \phi \right) }{\omega _{\phi }-3}.  \label{F20}
\end{equation}%
The DH function $f_{DH}\left( \phi \right) $ describes the PBH fraction if
and only if $\omega _{\phi }>3$. The formation of galaxies is mainly due to
the PBHs. In this scenario, the effective density Eq. (\ref{r8}) could be
the radiation density due to the PBH evaporation as the Hawking radiation.
There is a way to relate the PBHs fraction to the GB coupling, by setting $%
\omega _{\phi }=3D/4$. In the present context, we assume that there is a
continuity between $f\left( \phi \right) $ and $\alpha $ in galaxy\ edges,
Eq.(\ref{F11}). We can express the PBH fraction as follows $%
f_{PBH}=4f_{DH}\left( \phi \right) /3\left( D-4\right) $. By analogy, we
observe that $f\left( \phi \right) \equiv f_{PBH}\ $and $\alpha \equiv \frac{%
4}{3}f_{DH}\left( \phi \right) $. In the galaxy interior ($\rho _{DM}\approx
0$, \ $P_{\phi }\approx 0$) so we get $f\left( \phi \right) \approx 3\rho
_{\phi }/4\rho _{m}$. In this case, the relation (\ref{F20}) is not valid
only in the edges of the galaxy.

\section{Conclusion}

In summary, we have studied the model describing the rotation curves of
galaxies surrounded by scalar dark matter in 4D Einstein-Gauss-Bonnet
gravity. We have made a comparison between the galaxy interior and its
edges. In this case, the Gauss-Bonnet coupling describes the interior
structure of the galaxy, while the coupling function $f\left( \phi \right) $
describes the galaxy edges. Under this premise and within the framework of
the singlet scalar dark matter (DM) model, we have explored the impact of
primordial black holes on the galaxy's formation at early times. The flat
galactic rotation curves can be explained either by introducing the
Gauss-Bonnet coupling. We show that the difference between the rotation
curves of spiral and dwarf galaxies due to the effective mass varied from
one type of galaxy to another. So, this mass can determine the type of a
galaxy if we know its rotational velocity. The predictions for the galaxy
rotation curves from observations and our model agree remarkably for almost
all of the 35 galaxies. The effective mass range\ in our model is $%
10^{-2}kg-10^{3}kg$, which is in good agreement with the constraints on
primordial black holes, which shows that the mass hidden in the dark matter
halos is a mass of the primordial black holes. This opens a new window on
the galaxy's formation in the early Universe by PBHs production.

\end{document}